\documentclass{PoS}

\title{Scaling behavior in two-flavor QCD, finite quark masses and finite volume effects}

\ShortTitle{Scaling behavior, finite quark masses and finite volume}

\author{\speaker{Bertram Klein}\\
        Physik Department, Technische Universit\"at M\"unchen\\
        E-mail: \email{bklein@ph.tum.de}}

\author{Jens Braun\\
        Theoretisch-Physikalisches Institut, Friedrich-Schiller-Universit\"at Jena\\
        E-mail: \email{j.braun@uni-jena.de}}
\author{Piotr Piasecki\\
Insititut f\"ur Kernphysik, Technische Universit\"at Darmstadt\\
E-mail: \email{piotr@crunch.ikp.physik.tu-darmstadt.de}        
        }

\abstract{The question of the exact nature of the phase transition in two-flavor QCD is still under discussion. Recent results for small quark masses in simulations with $2+1$ flavors show scaling behavior consistent with the O(4) or O(2) universality class. For a precise determination, an assessment of deviations from the ideal scaling behavior due to finite quark masses and finite simulation volumes is necessary. \\

We study the scaling behavior at the chiral phase transition with an effective quark-meson model. 
In our Renormalization Group approach, the quark masses in the model can be varied from the chiral limit over a wide range of values, which allows us to estimate scaling deviations due to large quark masses and the extent of the scaling region. We conclude that scaling deviations are already large at pion masses of $75$ MeV, but that the effect is difficult to see in the absence of results for even smaller masses. Comparing results only in a narrow window of pion masses leads to the observation of apparent scaling behavior. 
While the scaling deviations are not necessarily universal, we expect that this may affect current lattice simulation results.
\\

By placing the system in a finite box, we investigate the transition between infinite-volume scaling behavior and finite-size scaling. We estimate that finite-size scaling behavior can be tested in regions where $M_\pi L \approx 2 - 3$, which is smaller than in most current lattice simulations. We expect that finite-volume effects are small for pion masses of $M_\pi = 75$ MeV and lattice aspect ratios with $TL \ge 8$,  but that they will become significant when pion masses in lattice simulations become smaller.}

\FullConference{ The XXIX International Symposium on Lattice Field Theory - Lattice 2011\\
July 10-16, 2011\\
Squaw Valley, Lake Tahoe, California}

\begin{document}

\section{Introduction}
The order and the exact nature of the chiral phase transition in QCD are currently being explored in  lattice simulations \cite{Aoki:2006we} which use critical scaling behavior for the analysis \cite{Ejiri:2009ac,Kaczmarek:2011zz}. For $N_{\mathrm{f}} = 2$ massless quark flavors, a second-order phase transition in the O(4) universality class is expected, and the more realistic case of $N_{\mathrm{f}}=2+1$ flavors with two light and one heavy flavor is likely also governed by this critical behavior. 
There have been numerous investigations of critical behavior and scaling in the O(4) universality class \cite{Engels:2001bq,Engels:2003nq,ParisenToldin:2003hq,Berges:1997eu,Schaefer:1999em,Bohr:2000gp,Litim:2001hk,Litim:2002cf,Braun:2007td,Braun:2008sg,Stokic:2009uv}, in spin model systems as well as in effective models for chiral symmetry breaking. Most of these studies have focussed on exploring the actual critical scaling regime. This requires to effectively go to very small masses for the Goldstone modes.

The idea of our study \cite{Braun:2010vd} is to use a simple model for the chiral phase transition to investigate the influence of a finite quark mass or a finite volume on the scaling behavior. The model shows the same universal behavior which is expected in QCD in the phase transition region. In the model, it is easy to obtain results both in the infinite-volume limit in the scaling region and for large pion masses and small volumes, where deviations from the scaling behavior become apparent. As far as these deviations are mainly determined by the long-range fluctuations, the results are applicable to the universal behavior at the chiral transition in QCD. In this way they can shed light on the scaling deviations one should expect  in the analysis of QCD lattice simulations. 

In the following, we will first present the model that we are using and discuss the scaling behavior in the ideal case of small pion masses and large volumes. We then turn to deviations from this scaling behavior in a finite volume. For very small volumes compared to the scale set by the pion mass, finite size scaling can be used for the analysis.  

\section{The quark-meson model}

The quark-meson model belongs to the family of Nambu--Jona-Lasinio type models and can be used to describe the dynamical breaking of the chiral $SU(N_{\mathrm{f}})_{L}\times SU(N_{\mathrm{f}})_{R}$ symmetry. We use the model with $N_{\mathrm{f}}=2$ light flavors, where the chiral symmetry is implemented as an O(4) symmetry in the meson sector. It does not include any gluonic degrees of freedom and is not confining for the quarks. 
We define the model at an ultraviolet scale $\Lambda$ by its bare effective action
\begin{eqnarray} 
  \Gamma_{\Lambda}[\bar \Psi,\Psi,\phi]&=& \int d^{4}x \Big\{
  \bar{\Psi} \left({\rm i}{\partial}\!\!\!\!\!\slash + 
  g(\sigma+i\vec{\tau}\cdot\vec{\pi}\gamma_{5})\right)\Psi
  +\frac{1}{2}(\partial_{\mu}\phi)^{2}+U_{\Lambda}(\phi^2) - H \sigma \Big\} 
\label{eq:QM}
\end{eqnarray} 
with an effective potential specified by two parameters. In the ground state, the symmetry is dynamically broken, and the first component of $\phi=(\sigma, \vec{\pi})$ acquires a finite expectation value. There are $3$ Goldstone bosons associated with the remaining O(3) symmetry of the ground state.
The symmetry-breaking parameter $H$ controls the amount of explicit symmetry breaking and the mass of the Goldstone bosons. With our choice of parameters, the model undergoes a phase transition at $T_c\approx 145$ MeV to a phase with restored symmetry. 

In order to account for the critical fluctuations at the phase transition, which determine the behavior we want to study, we use a non-perturbative Renormalization Group (RG) method and solve the Wetterich equation \cite{Wetterich:1992yh} for the RG flow in the local potential approximation.
For details of our model and calculation, we refer the reader to \cite{Braun:2010vd}.

\begin{figure}[t]
\hspace{-8mm}\includegraphics[scale=0.65]{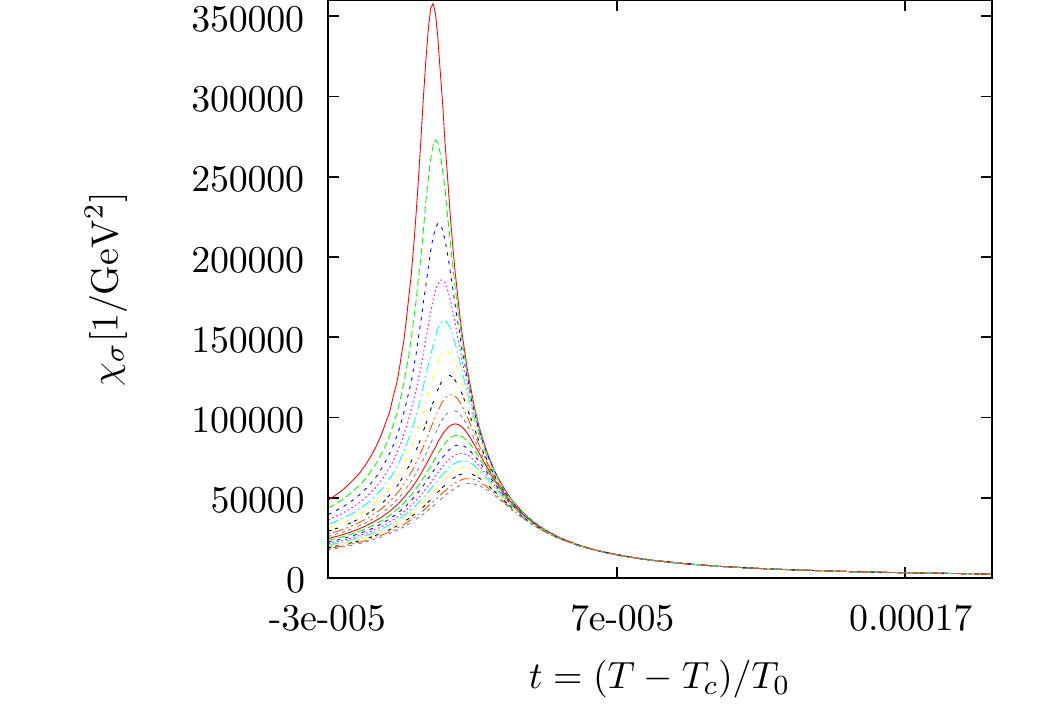}
\includegraphics[scale=0.65]{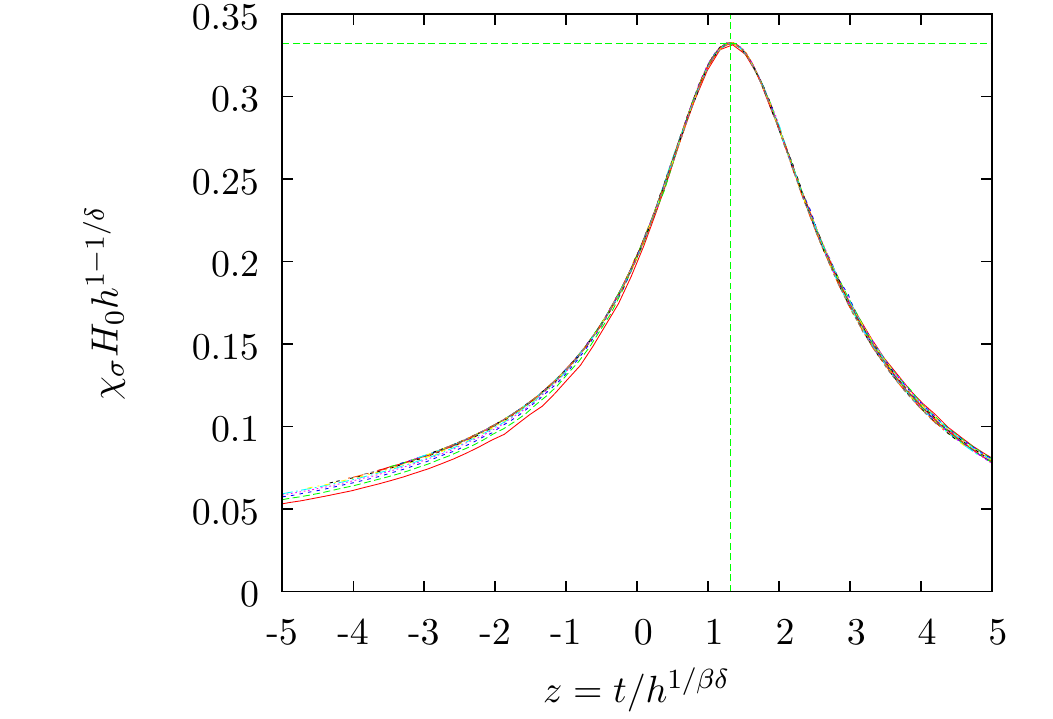}
\caption{Left panel:  Longitudinal suszeptibility $\chi_\sigma$ as a function of temperature $t$ for various pion masses $M_\pi< 0.9$ MeV.
Right panel: Rescaled susceptibility as a function of $z$ for the same pion masses. For these small pion masses, all curves fall onto the scaling function with good accuracy.}
\label{fig:smallmpi}
\end{figure}

Because of the effect of critical fluctuations, close to the critical point the behavior of thermodynamic quantities can be expressed in terms of power laws with universal critical exponents which depend on symmetry and dimensionality of the system. As an additional consequence, the dependence on temperature $T$ and symmetry-breaking parameter $H$ can be expressed in terms of a scaling function of a single scaling variable $z$. In the case of the order parameter $M$ (the chiral condensate or pion decay constant $f_\pi$ in our case), this takes the form
\begin{eqnarray}
M(t, h) = h^{1/\delta} f_M(z), \;\; z = t/h^{1/(\beta\delta)}. 
\end{eqnarray}  
In order to be able to compare different systems in the same universality class, one defines the universal quantities $t=(T-T_c)/T_0$ and $h=H/H_0$, where the normalization constants $T_0$ and $H_0$ are non-universal and like the critical temperature $T_c$ system-specific. The corresponding relation for the longitudinal susceptibility is given by
\begin{eqnarray}
\chi_\sigma &=& \frac{h^{1/\delta-1}}{H_0} f_\chi(z) = \frac{h^{1/\delta-1}}{H_0} \frac{1}{\delta}  \left[ f_M(z) -\frac{z}{\beta} f_M^\prime(z)\right].
\end{eqnarray}
We have checked carefully that our results for very small pion masses ($M_\pi < 1$ MeV) satisfy these scaling relations to a very high accuracy. In particular for the susceptibility it is impressive how well the results for different pion masses over several orders of magnitude collapse onto the scaling function after rescaling, see Fig.~\ref{fig:smallmpi} and Ref.~ \cite{Braun:2010vd}.

\section{Scaling for large quark masses}

\begin{figure}[t]
\hspace{-10mm}\includegraphics[scale=0.68]{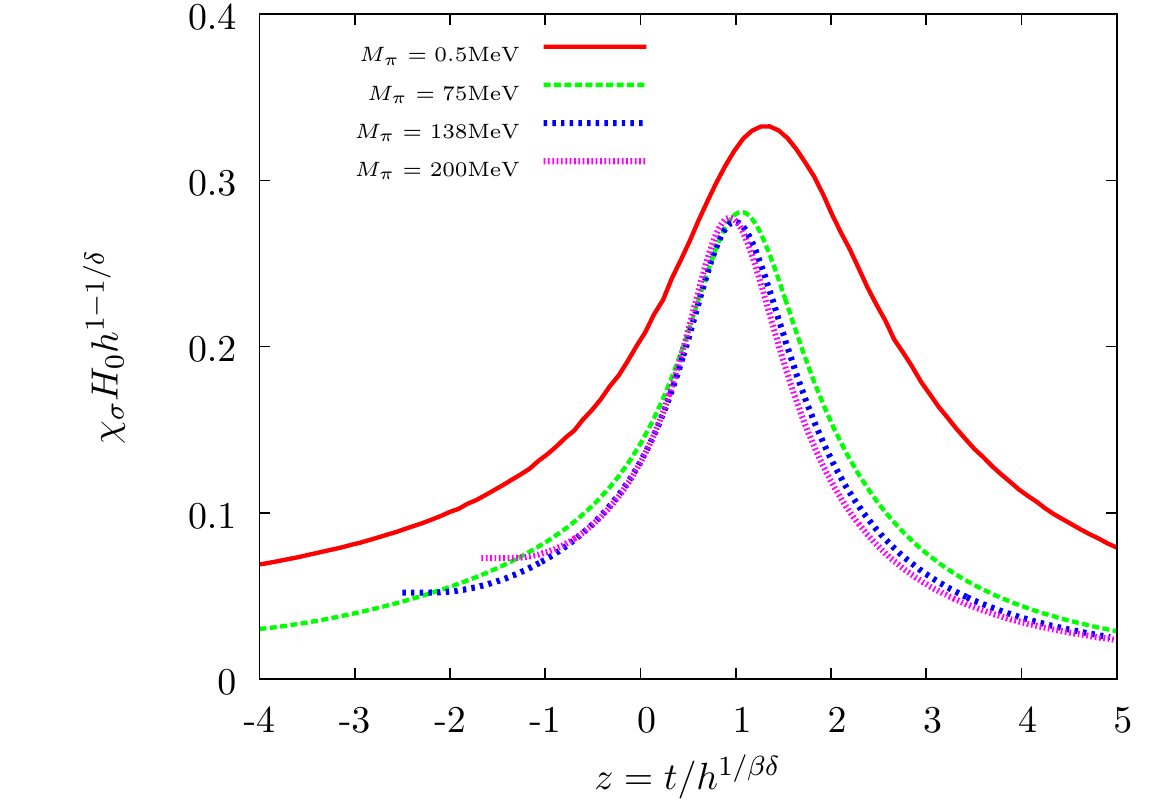}
\includegraphics[scale=0.68]{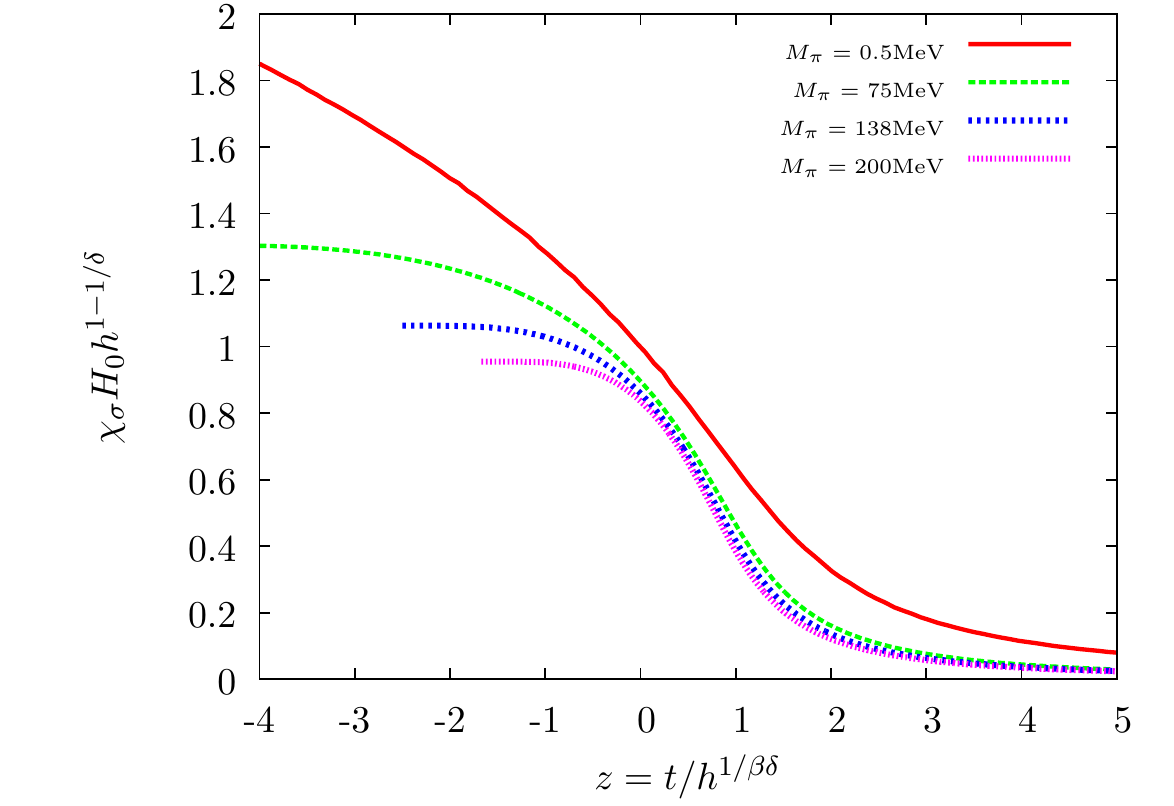}
\caption{Left panel: Rescaled $\sigma$-suszeptibility as a function of $z$ for various pion masses.
Right panel: Rescaled pion decay constant $f_{\pi}$ (order parameter) as a function of $z$ for various pion masses. In the absence of the results for small pion masses, the curves for large pion masses appear to fall on the scaling function.}
\label{fig:largempi}
\end{figure}
While the expected intinite-volume scaling behavior is realized impressively well in the model for a small amount of symmetry breaking, we find that it is necessary to remain close to the critical point in order to    
keep the scaling deviations small. An important question is how large these deviations are for realistic values of the pion mass and for pion masses used in lattice simulations. Scaling deviations are of course not universal and can reflect the different short-range physics of the model compared to QCD. However, the absolute scales (transition temperature) for the effective degrees of freedom at the transition are very similar, and therefore we expect similar deviations, provided the behavior is indeed mainly governed by the low-energy degrees of freedom. 

In Fig.~\ref{fig:largempi} we show results for the rescaled susceptibility and the rescaled order parameter as a function of the scaling variable $z$ for pion masses of $M_\pi = 75, 138, 200$ MeV.  We find that the rescaled results differ significantly from the scaling function that we obtained in the limit of small pion masses. The resulting scaling functions are suppressed compared to the proper one, and the position of the peak in the susceptibility is shifted compared to the small-pion mass limit. 

In the absence of the results for small pion masses, the rescaled results obtained for large pion mass would appear to agree quite well and to actually show scaling behavior: The curves are almost on top of one another and differ significantly only far from the critical temperature, below the transition. Using these results to determine new values for the normalization constants $T_0$ and $H_0$, they can be rescaled to almost agree with the original scaling function. 

We conclude that the scaling deviations at pion masses of $M_\pi = 75$ MeV are already quite large, for both the longitudinal susceptibility and the order parameter, but that this effect is difficult to observe in the absence of results in the limit of very small symmetry breaking. Comparing only results in a narrow window of large pion masses leads to the observation of apparent scaling, while these results are still very different from the correct scaling function.   

\section{Scaling in a finite volume}

The RG calculation can easily be continued to a finite volume $V=L^3$  \cite{Braun:2004yk,Braun:2005gy,Braun:2005fj,Braun:2008sg}. In a finite volume, the boundary conditions in the spatial direction for quark fields can be chosen freely; we chose periodic boundary conditions, which is the choice taken in most QCD lattice simulations. Our main interest in this part of the study was to find out how much the results for the scaling behavior change due to the finite volume if we performed the same scaling analysis as in infinite volume, i.e determined the scaling function $f_\chi(z)$ for the susceptibility.  

\begin{figure}[t]
\hspace{-10mm}
\includegraphics[scale=0.64]{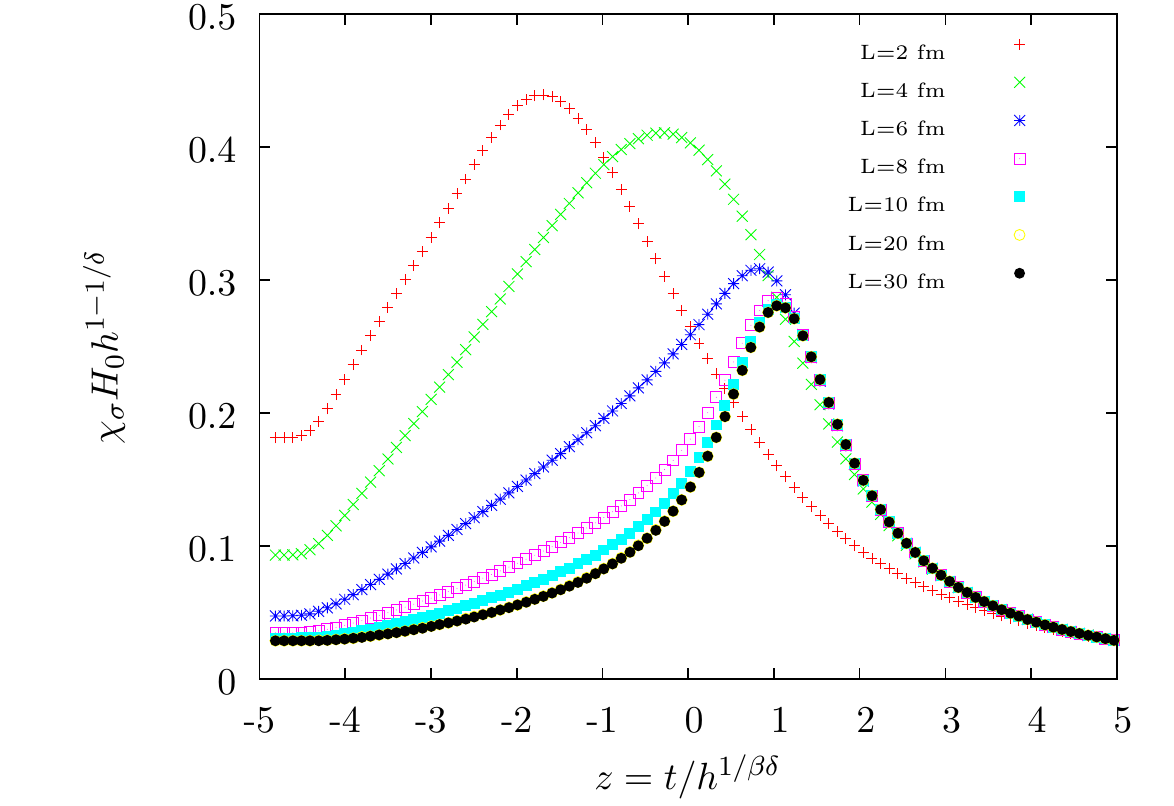}
\includegraphics[scale=0.72]{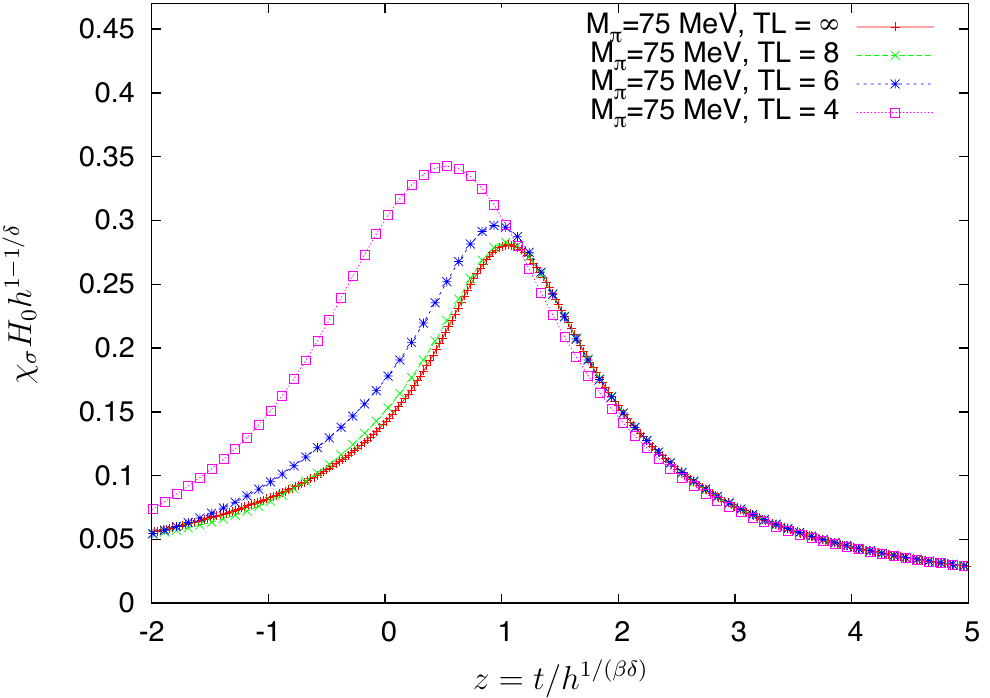}
\caption{Rescaled longitudinal susceptibility $\chi_\sigma$ for $M_{\pi}= 75$ MeV as a function of $z$ for various box lengths $L$ (left panel), and for a fixed ratio $TL$ of box size $L$ and Euclidean time extent $1/T$.}
\label{fig:susc_finiteL}
\end{figure}

The results are shown in Fig.~\ref{fig:susc_finiteL} for a pion mass $M_\pi = 75$ MeV. In the left panel, we show the results for the rescaled susceptibility $ h^{1/\delta-1 } H_0 \chi_\sigma$ as a function of $z$ for several different volume sizes. Significant deviations occur only for volume sizes smaller than $L =8$ fm. However, in usual lattice simulations at finite temperature, the aspect ratio of the spatial and temporal directions of the Euclidean volume is kept fixed and both volume size and temperature are varied at the same time. In addition to the results already given in Ref.~\cite{Braun:2010vd}, we show in the right panel of Fig.~\ref{fig:susc_finiteL} results for this scenario, where the aspect ratio $TL$ of the Euclidean volume has been kept fixed. We find that for this pion mass, deviations appear only below $TL=6$. We conclude from these results that finite-volume effects presumably do not play a large role in current lattice simulations of the scaling behavior. However, they are significant and likely to become important for simulations at smaller pion masses. 

A somewhat surprising result is the {\it increase} in the susceptibility in an intermediate volume range that can be seen in Fig.~\ref{fig:susc_finiteL}, which we attribute partially to quark effects \cite{Braun:2005gy}. Naively one expects $\chi_\sigma \sim L^2$, but this behavior is only realized for very small volumes. These finite-volume effects are highly dependent on the boundary conditions chosen for the quark fields \cite{Braun:2005gy} and can also influence the response of the system to a finite chemical potential \cite{Klein:2010tk,Braun:2011iz}.

\section{Finite-size scaling}

For very small volume size, when the correlation length of the critical fluctuations is of the same order as the volume extent $L$, scaling with the volume size can be observed. Starting from the hypothesis that the system behavior should be unaltered when both system size $L$ and correlation length $\xi$ are changed such that $\xi/L$ remains constant, it is possible to derive scaling functions for the volume dependence \cite{Engels:2001bq,Braun:2008sg}. These scaling functions will depend on both the infinite-volume scaling variable $z$ and a new scaling variable involving $L$. For example, for the longitudinal susceptibility one finds
\begin{eqnarray}
\chi_\sigma(t, h, L) &=& L^{\gamma/\nu} Q_\chi(z, hL^{\beta \delta/\nu}) 
\end{eqnarray}
The new finite-size scaling function $Q_\chi$ satisfies the constraint that the infinite-volume scaling behavior must be recovered in the infinite-volume limit. 

\begin{figure}
\hspace{-5mm}
\includegraphics[scale=0.68]{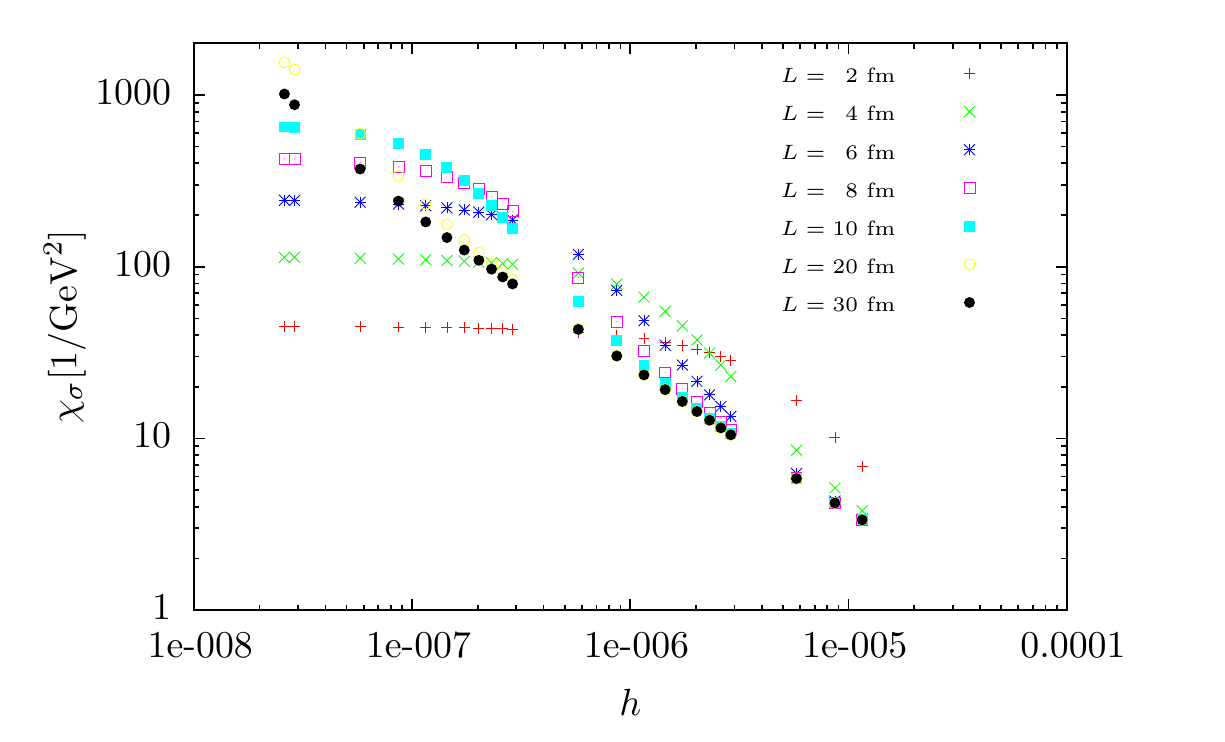}
\hspace{-8mm}\includegraphics[scale=0.68]{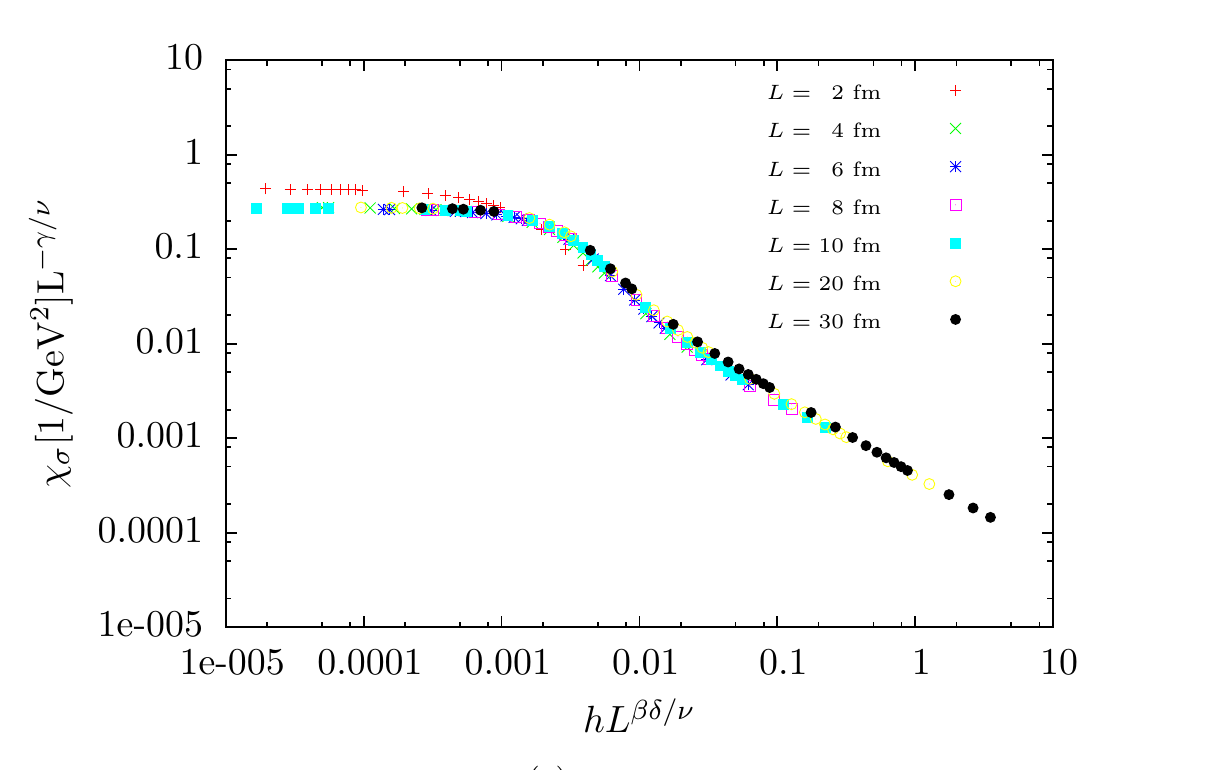}
\caption{Finite-size scaling for the longitudinal susceptibility $\chi_{\sigma}$ for $z=0$. The unscaled susceptibility is shown as a function of the symmetry breaking parameter $h$ (left panel) and the rescaled susceptibility $\chi_\sigma L^{-\gamma/\nu}$ is shown as a function of the scaling variable $hL^{\beta\delta/\nu}$ for various values of the box length~$L$.}
\label{fig:fss}
\end{figure}

This scaling behavior for the longitudinal susceptibility is shown in Fig.~\ref{fig:fss} for $T=T_c$. In this case the symmetry-breaking parameter $h$ controls the correlation length, an increase in $h$ translates to larger masses of the fluctuations and smaller correlation length. In the left panel, it is evident that the susceptibility deviates from the asymptotic infinite-volume behavior at different $h$-values, depending on the volume size. The right panel shows the rescaled susceptibilities, which all fall onto the same scaling function. 

Since the volume size is effectively an additional relevant coupling controlling the critical behavior, $L$ needs to be renormalized with a non-universal factor $L_0$, just as $T$ and $H$, in order to facilitate a comparison of different systems. In the present case, it appears more useful to translate the scale into terms of the dimensionless quantity $M_\pi L$. The values of this quantity at the bend, where the scaling function for $\chi_\sigma$becomes flat, are given in Tab.~\ref{tab:bend}. In order to observe finite-size scaling effects and to use finite-size scaling as a tool for the analysis, it appears necessary to explore the region in which $M_\pi L \approx 2 - 3$. This is much smaller than the value achieved in most present lattice simulations, where generally $M_\pi L > 4 - 5$ and finite-volume effects are minimized. For a systematic exploration of finite-size scaling effects, {\it smaller} lattice sizes would be required. 
\begin{table}
\begin{center}
\begin{tabular}{lrrrrrrr}
\hline\hline
$L$ [fm] & 
\phantom{0}2 & 
\phantom{0}4 & 
\phantom{0}6 &
\phantom{0}8 &
10 &
20 &
30 \\\hline
$M_\pi$ [MeV] &
\phantom{X}308&
\phantom{X}139&
\phantom{XX}85&
\phantom{XX}60&
\phantom{XX}45&
\phantom{XX}19&
\phantom{XX}11\\
\hline
$M_\pi L$ & 
3.12&
2.82&
2.59&
2.43&
2.30&
1.94&
1.75\\
\hline
\hline
\end{tabular}
\end{center}
\caption{Infinite-volume pion masses $M_\pi(T\to 0, L\to \infty)$ which correspond to the value of the scaling variable $hL^{\beta\delta/\nu}$  at the point where the scaling curve in Fig.~\protect\ref{fig:fss} 
bends (for $z=0$). The dimensionless product $M_\pi L$ contains only long-range quantities and thus provides a benchmark for comparison to QCD simulations.}
\label{tab:bend}
\end{table}

\section{Conclusions}
We have presented results from a scaling analysis of the chiral phase transition in a quark-meson model. In such a model the scaling region at small pion masses can be explored as well as deviations from this behavior for large pion masses and finite volumes. In order to properly account for critical fluctuations, we used a non-perturbative RG method for the calculation. As a main result, we find that for $M_\pi \gtrsim 75$ MeV significant deviations appear. This suggests that such scaling deviations could also be significant in current lattice QCD simulations. In contrast to the effects of finite pion masses, we find finite-volume effects to be small for current simulations parameters. In order to use finite-size scaling for the analysis of the phase transition behavior, it is necessary to go to much smaller volumes for current pion mass values, such that $M_\pi L = 2 - 3$.

\section*{Acknowledgements}
BK acknowledges support by the DFG Research Cluster "Structure and Origin of the Universe". JB acknowledges financial support by the DFG under Grant BR 4005/2-1 and the DFG 
research training group GRK 1523/1.

\end{document}